\documentclass{PoS}
\PoS{PoS(LAT2005)291}

\usepackage{epsfig,multicol}

\newcommand\fverb{\setbox\pippobox=\hbox\bgroup\verb}
\newcommand\fverbdo{\egroup\medskip\noindent%
			\fbox{\unhbox\pippobox}\ }
\newcommand\fverbit{\egroup\item[\fbox{\unhbox\pippobox}]}
\newcommand{\Slash}[1]{\ooalign{\hfil/\hfil\crcr$#1$}}
\newbox\pippobox

\def\NPB{{Nucl. Phys.} {\bf B}}
\def\PLB{{Phys. Lett.} B}

\def\PRD{{Phys. Rev.} D}

\title{The running coupling in lattice Landau gauge\\
with unquenched Wilson fermion and KS fermion}

\author{\speaker{Sadataka Furui} and Hideo Nakajima$^\dagger$\\
	$^*$ School of Sci.\& Engr., Teikyo Univ., Utsunomiya, 320-8551 Japan\\
	E-mail: \email{furui@umb.teikyo-u.ac.jp}\\
	$^\dagger$ Dept. of Infor. Sci., Utsunomiya Univ., Utsunomiya, 320-8585 Japan\\
	E-mail: \email{nakajima@is.utsunomiya-u.ac.jp}}

\abstract{The running coupling of the Wilson fermon(JLQCD/CP-PACS) and
that of Kogut-Susskind(KS) fermion(MILC) are measured in the lattice Landau gauge QCD in $\widetilde{MOM}$ scheme. The quark propagator of the KS fermion is
also measured and we find that it is infrared suppressed.  The renormalization factor of the running coupling and the tadpole renormalization define the scale of the quark wave function.
Effects of the $A_\mu^2$ condensates of a few GeV$^2$ are observed in the running coupling and also in the quark propagator. }

\FullConference{XXIIIrd International Symposium on Lattice Field Theory\\
25-30 July 2005\\
Trinity College, Dublin, Ireland}
\ShortTitle{The running coupling in lattice Landau gauge}

\begin{document} 


\section{Introduction}

The mechanism of dynamical chiral symmetry breaking and confinement is one of the most fundamental problem of hadron physics. The propagator of dynamical quarks in the infrared region provides information on dynamical chiral symmetry breaking and confinement.  In the previous paper\cite{FN04}, we measured gluon propagators and ghost propagators of unquenched gauge configurations obtained with quark actions of Wilson fermions (JLQCD/CP-PACS) and those of Kogut-Susskind(KS) fermions (MILC) \cite{MILC1,MILC2} in Landau gauge and observed that the configurations of the KS fermion are closer to the chiral limit than those of Wilson fermions.  

In the analysis of running coupling obtained from the gluon propagator and the ghost propagator, with use of the operator product expansion of the Green function, we observed possible contribution of the quark condensates and $A^2$ condensates in the configurations of the KS fermion\cite{FN04}. The quark propagator of quenched KS fermion was already measured in \cite{ABB}, and possible contribution of these condensates are reported. Unquenched KS fermion propagator of $20^3\times 64$ lattice (MILC$_c$) was measured in \cite{bhlpwz}, but to distinguish the gluon condensates and the quark condensates, it is desirable to measure the quark propagator of larger lattice (MILC$_f$) and to compare with data of MILC$_c$.
We measure quark propagator of gauge configuration of 1) MILC$_c$ $20^3\times 64$, $\beta=6.76$ and 6.83 and 2) MILC$_f$ $28^3\times 96$, $\beta=7.09$ and 7.11, using the Staple+Naik action\cite{OT}.

\section{The running coupling in $\widetilde{MOM}$ scheme}

The running coupling in $\widetilde{MOM}$ scheme is given with use of the vertex renormalization factor $\tilde Z_1$ as
\begin{equation}\label{alphadef}
\alpha_s(q)=\alpha_R(\mu^2)Z_R(q^2,\mu^2){G_R}(q^2,\mu^2)^2\nonumber\\
=\frac{\alpha_0(\Lambda_{UV})}{{\tilde Z_1(\beta,\mu)}^2} Z(q^2,\beta){G}(q^2,\beta)^2
\end{equation}
where $Z$ and $G$ are the gluon and the ghost dressing function, respectively, and $\tilde Z_1$ is the vertex renormalization factor.
\smallskip
\DOUBLEFIGURE[b]{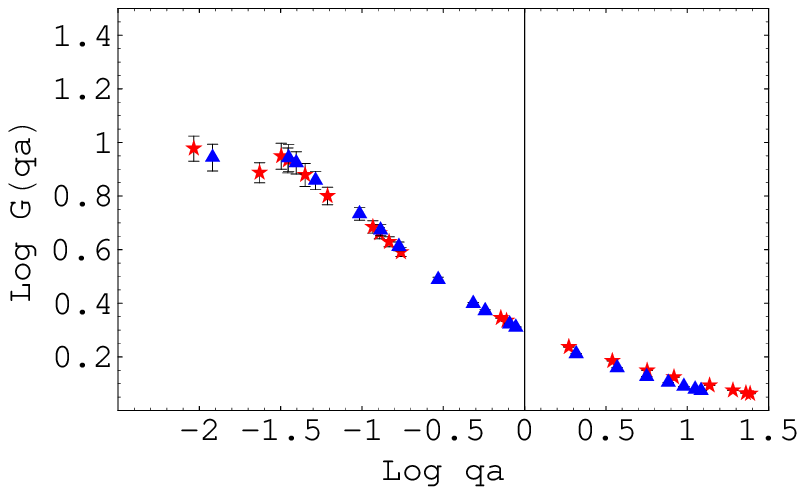,width=7cm} {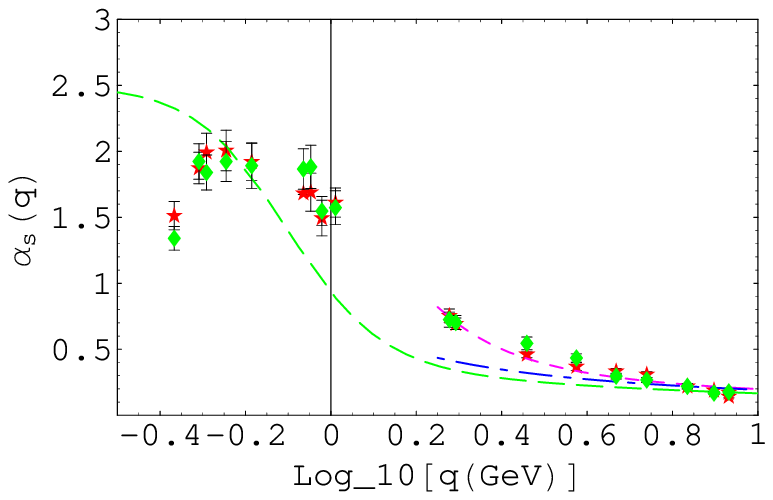,width=7cm} {The $\log G(qa)$ as a function of  $\log qa$ ($a$ is the lattice spacing of MILC$_f$) of MILC$_f$ $\beta_{imp}=7.09$(stars) and MILC$_c$ $\beta_{imp}=6.76$(triangles).} {The running coupling $\alpha_s(q)$ as a function of $\log_{10}q$(GeV) of MILC$_f$  $\beta_{imp}=7.09$(stars) and 7.11(diamonds).}\label{alp709_711}

Our data suggests that the gluon propagator is infrared finite as in Dyson-Schwinger equation\cite{Blo1}. The running coupling in the infrared is suppressed as shown in Figure 2,
but the main origin is the suppression of the ghost propagator in the infrared.

We parametrize the difference of the lattice data and the pQCD 4-loop result\cite{ChRe,chet} in the 1GeV$<q<6$GeV region in the form, with a minor correction term $d$ as
\begin{equation}
 \Delta \alpha_s(q)=\alpha_{s,latt}(q)-\alpha_{s,pert}(q)=\frac{c_1}{q^2}+\frac{c_2}{q^4}+d,
\end{equation}
where the $A^2$ condensates gives $c_1$ and the gluon condensates and/or quark condensates gives $c_2$. Although statistics is not large, running coupling of CP-PACS suggests $c_1\sim 2$GeV. The MILC data suggests $c_1\sim 4$GeV and  $c_2\sim -2$GeV. There is an analytical calculation that suggests correlation between the $A_2$ condensates and the the horizon function parameter\cite{dssv}.

\section{The quark wave function renormalization}

We renormalize the quark field as $\psi_{bare}=\sqrt{Z_2}\psi_R$, 
and define the colorless vector current vertex\cite{Orsay1}
\begin{equation}
\Gamma_\mu(q,p)=S^{-1}(q)G_\mu(q,p)S^{-1}(q+p)
\end{equation}
where
\begin{equation}
G_\mu(p,q)=\int d^4 x d^4 ye^{ip\cdot y+iq\cdot x}\langle q(y)\bar q(x)\gamma_\mu q(x)\bar q(0)\rangle
\end{equation}
and $S(q)$ is the quark propagator.

The vertex of the vector current with $p=0$ is written as
\begin{equation}
\Gamma_\mu(q)=\delta_{a,b}\{g_1(q^2)\gamma_\mu+ig_2(q^2)p_\mu+g_3(q^2)q_\mu\Slash{q}+ig_4(q^2)[\gamma_\mu,\Slash{q}]\}
\end{equation}
The Ward identity implies
\begin{equation}
Z_V^{\widetilde{MOM}}\Gamma_\mu (q)=-i\frac{\partial}{\partial q^\mu} S^{-1}(q)
\end{equation}
where $Z_V^{\widetilde{MOM}}g_1(q^2)=Z_\psi(q^2)$ and $Z_V^{\widetilde{MOM}}=1$ when there is no lattice artefact.

The running coupling $g$ of the ghost, anti-ghost, gluon coupling 
\begin{equation}
g(q)=\tilde Z_1^{-1} Z_3^{1/2}(\mu^2,q^2) {\tilde Z}_3(\mu^2,q^2)
\end{equation}
and that of quark, gluon coupling 
\begin{equation}
g(q)=Z_1^{-1} Z_3^{1/2}(\mu^2,q^2) Z_2(\mu^2,q^2)
\end{equation}
are identical due to the Slavnov-Taylor identity.
At the renormalization point $q=\mu$,  we fix $Z_2(\mu^2,\mu^2)=1$
and $\tilde Z_3(\mu^2,\mu^2)=1$, and so 
$\tilde Z_1^{-1} Z_3^{1/2}{\tilde Z}_3=Z_1^{-1} Z_3^{1/2} Z_2$
implies $\tilde Z_1=Z_1$. On the other hand
$Z_\psi(q^2)=Z_V^{\widetilde{MOM}}g_1(q^2)$
where for 16 tastes
\begin{equation} g_1(q^2)=\frac{1}{48N_c}Tr[\Gamma_\mu(q,p=0)(\gamma_\mu-q_\mu\frac{\Slash{q}}{q^2})]
\end{equation}
When $Z_V^{\widetilde{MOM}}=1$, $g_1(\mu^2)$ is identical to $\tilde Z_1$ which is defined by the renormalization of the running coupling on the lattice defined 
at $\mu\sim 6$GeV and summarized in Table \ref{z1fac}.

\begin{table}[htb]
\begin{center}
\begin{tabular}{c|c|c|c|c}
         &  $\beta_1/K_{sea 1}$ & $\beta_2/K_{sea 2}$ & average & configurations\\
\hline
CP-PACS  & 1.07(8) & 1.21(10) & 1.14 & $K_{sea 1,2}=0.1357,0.1382$\\
\hline
MILC$_c$ &  1.49(11) & 1.43(10)&  1.46 &$\beta_{1,2}=6.83, 6.76$\\
MILC$_f$ &  1.37(9)  & 1.41(12)&  1.40 &$\beta_{1,2}=7.11, 7.09$\\
\hline
\end{tabular}
\caption{The $1/\tilde Z_1^2$ factor of the unquenched SU(3).}\label{z1fac}
\end{center}
\end{table}
\smallskip
\DOUBLEFIGURE[t]{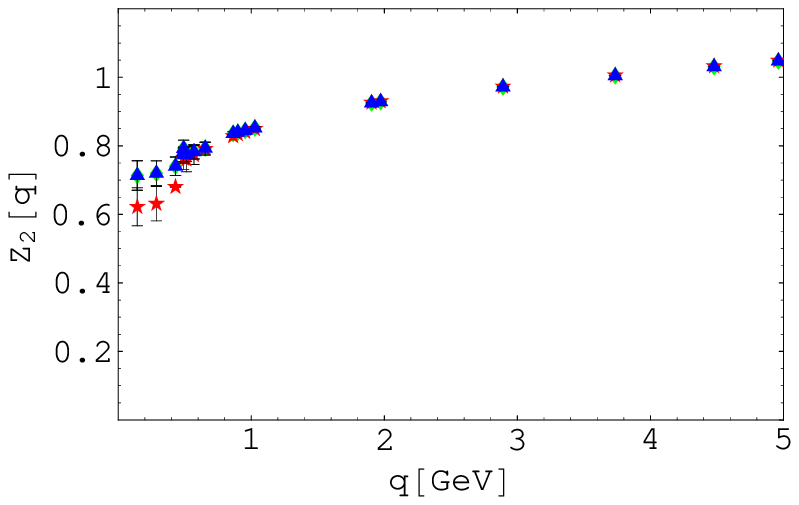,width=7cm} {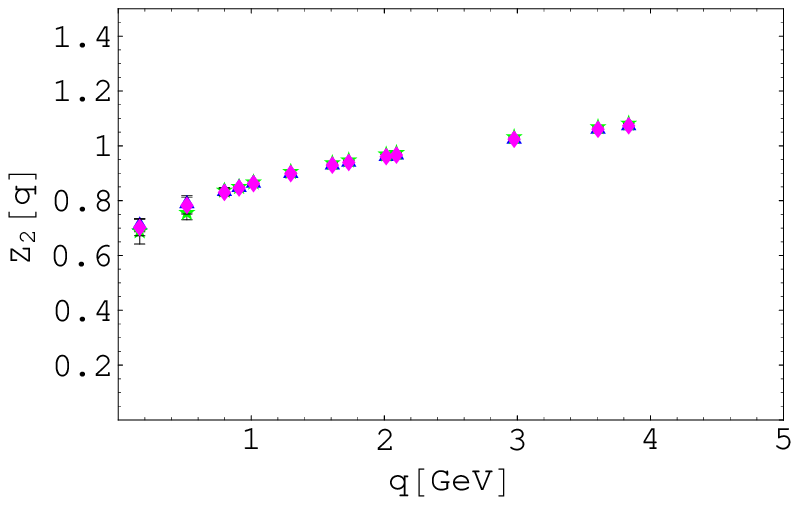,width=7cm} {The  renormalization factor $Z_2(q)$ of MILC$_f$ with bear mass $m_0=13.6$MeV(stars), 27.2MeV(diamonds) and 68.0MeV(triangles).} {Same as Fig. 3 but quark of MILC$_c$ with bear mass $m_0=11.5$MeV(stars), 65.7MeV(diamonds) and 82.2MeV(triangles). The last two almost overlap.}\label{z2_MILC}
\smallskip
\FIGURE[r]{\epsfig{file=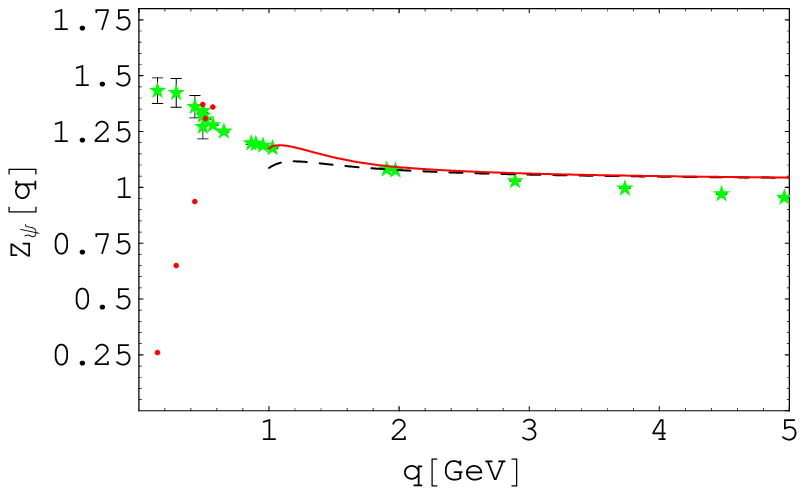,width=7cm} \caption{The quark wave function renormalization factor $Z_\psi(q)$ of MILC$_f$ with bear mass $m_0=27.2$MeV. Dashed line is the pQCD result and solid line is the result including the $A_\mu^2$ condensates. The points below 0.6GeV are $g_1(q^2)$ approximated by the running coupling obtained by the ghost anti-ghost gluon vertex.}\label{zpsi2896}}

The MILC configurations are produced by the Asqtad action with the bare masses 13.6MeV,

\noindent 27.2MeV  and 68.0MeV in the MILC$_f$, and 11.5MeV, 65.7MeV and 82.2MeV in the MILC$_c$.  The gluon wave function renormalization
$Z_3$ of the Asqtad action is renormalized by $1/u_0^2$, where $u_0$ is the fourth root of the plaquette value. The tadpole renormalization on quark wave function is $u_0$ and we normalize $S(q)$ by multiplying the average of $u_0\tilde Z_1$ i.e. 1/1.36 for MILC$_f$ and 1/1.38 for MILC$_c$. 
The quark propagators $Z_2(q)$ after this renormalization are infrared suppressed as shown in Figures 3 and 4. The apparent difference in the formulae of \cite{bhlpwz} and our work are only in the expression and in fact they are equivalent and the results agree with \cite{bhlpwz}.

Using the pQCD result of the inverse quark propagator\cite{ChRe,chet} and
 $\langle A^2\rangle$ and $\bar c_2$ (the contribution from the mixed condensate $\langle \bar q\Slash{A}q\rangle$\cite{LO}) as fitting parameters, we calculate 
\begin{equation}
Z_\psi(q)=\frac{1}{Z_2(q)}=Z_\psi^{pert}(q^2)+\frac{\left(\frac{\alpha(\mu)}{\alpha(q)}\right)^{{-\gamma_0+\gamma_{A^2}}\over{\beta_0}}}{q^2}\frac{\langle A^2(\mu)\rangle}{4(N_c^2-1)}{Z_\psi}^{pert}(\mu^2)+\frac{\bar c_2}{q^4}
\end{equation}
where $\alpha(q)$ are data calculated in the $\widetilde{MOM}$ scheme using the same MILC$_f$ gauge configurations\cite{FN04}. We estimate that $\bar c_2$ is small. In Figure 5, we show a fit of the MILC$_f$ $m_0=13.6$MeV data, by taking the renormalization point at $\mu\sim 3.8$GeV with use of $\langle A^2(\mu)\rangle\sim 1.6(3)$GeV$^2$ and $\bar c_2=0$. These parameters are consistent with \cite{Orsay1}.

In $q>1.5GeV$ region, dynamical mass of a quark in pQCD is expressed as\cite{ABB}
\begin{equation}\label{massf}
M(q)=-\frac{4\pi^2 d_M\langle \bar q q\rangle_\mu [\log (q^2/\Lambda_{QCD}^2)]^{d_M-1}}{3q^2 [\log (\mu^2/\Lambda_{QCD}^2)]^{d_M}}
+\frac{m(\mu^2)[\log (\mu^2/\Lambda_{QCD}^2)]^{d_M}}{[\log (q^2/\Lambda_{QCD}^2)]^{d_M}},
\end{equation}
where $d_M=\frac{12}{33-2N_f}$. The second term is the contribution of the massive quark. 
In the analysis of the lattice data, we observe that the quark condensates $-\langle \bar q q\rangle_\mu$ and $\Lambda_{QCD}$  roughly satisfy $-\langle \bar q q\rangle_\mu=(0.70\Lambda_{QCD})^3$\cite{Blo1}, with $\Lambda_{QCD}=0.69$GeV.

For the global fit of $M(q)$, we try the phenomenological monopole type\cite{SkW}
\begin{equation}
M(q)=\frac{c\Lambda^3}{q^2+\Lambda^2}+m_0
\end{equation} 
where $m_0$ is the bare quark mass.

Figures 6 and 7 show the mass function $M(0)$ of MILC$_f$ and MILC$_c$, respectively.

\smallskip
\DOUBLEFIGURE[h]{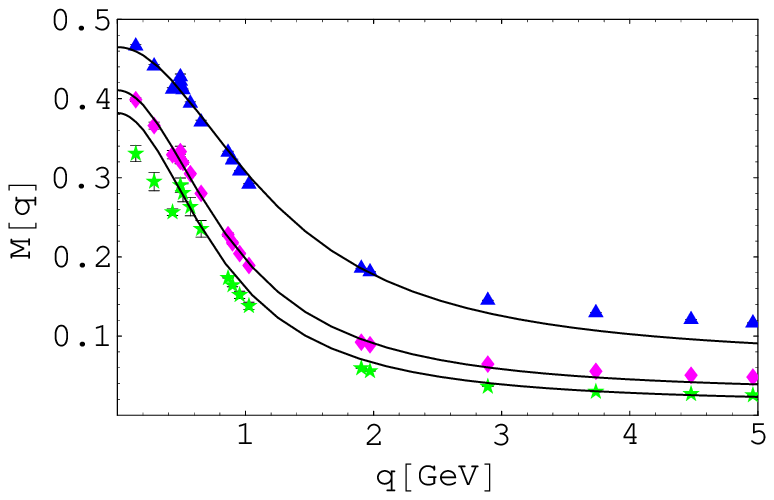,width=7cm} {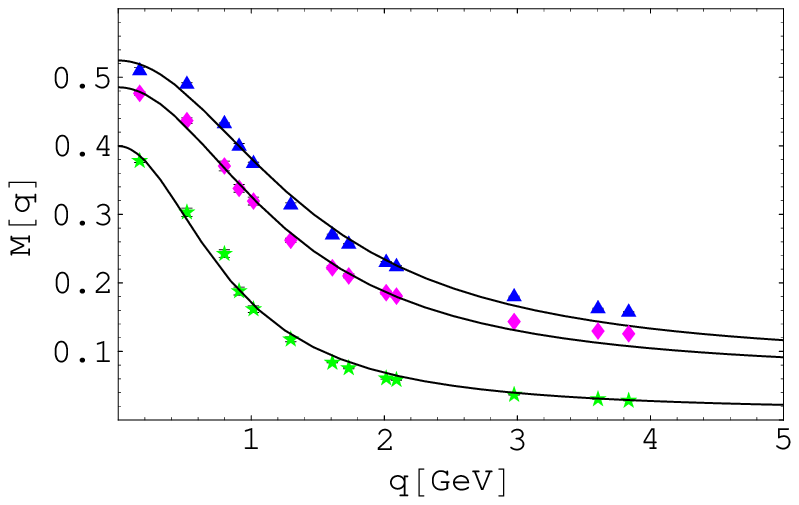,width=7cm}{The dynamical mass of the MILC$_f$ quark with bear mass $m_0=13.6$MeV(stars), 27.2MeV(diamonds) and 68.0MeV(triangles) and the phenomenological fits.} {The dynamical mass of the MILC$_c$ quark with bear mass $m_0=11.5$MeV(stars), 65.7MeV(diamonds) and 82.2MeV(triangles) and the phenomenological fits.}\label{massfunc}
\smallskip
\FIGURE[r]{\epsfig{file=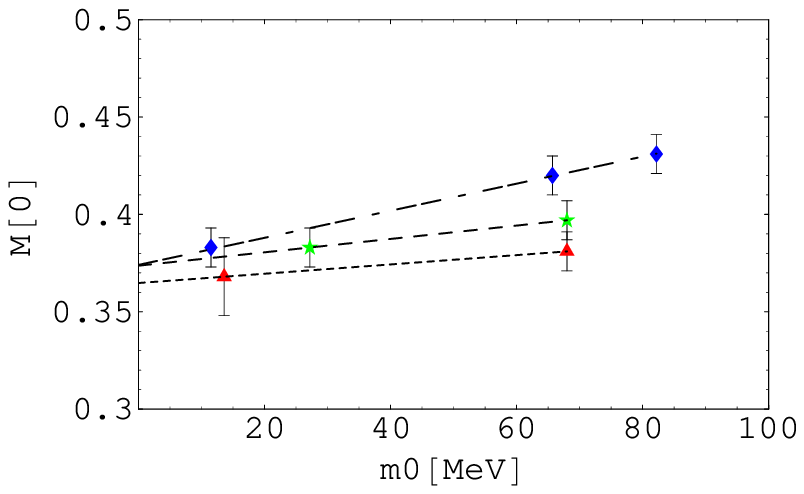,width=7cm} \caption{The mass function $M(0)$ as a function of bare mass. Dotted line is MILC$_f$ $\beta=7.09$, dashed line is $\beta=7.11$ and the dash-dotted line is MILC$_c$.}\label{chiralmass}}

We observe that the product  $c\Lambda$  becomes larger as the bare quark mass becomes heavy and it depends on $\beta$ in the case of MILC$_f$ but not in the case of MILC$_c$. In the case of MILC$_f$ $m_0=13.6$MeV,  the lowest three momentum points of $M(q)$ are systematically smaller than the other points. Ignoring these points we find $c\Lambda$ of $\beta=7.09$ is smaller than that of 7.11 and that of MILC$_c$ as shown in Figure \ref{chiralmass}. In the chiral limit $m_0\to 0$, we obtain $M(0)=0.35\sim 0.37$GeV, which is larger than \cite{bhlpwz} and that of the Wilson fermion\cite{SkW} by about 20\%.

\section{Discussion and conclusion}

We measured running coupling of unquenched Wilson fermion and KS fermion and the quark wave function renormalization factor and mass function of the KS fermion. The renormalization factor $Z_\psi$ obtained as 1/$Z_2$ is infrared finite. The Kugo-Ojima confinement criterion favours infrared vanishing of $Z_\psi$ and $g_1(q^2)$ approximated by the running coupling suggests this behavior, but it could be a lattice artefact. With infrared finite $Z_2$, infrared vanishing of $Z_{1\psi}$ is necessary for the confinement criterion to be satisfied.

\smallskip
\leftline{\bf Acknowledgement}
 We thank the MILC collaboration and JLQCD/

\noindent CP-PACS collaboration for supplying their gauge configurations in the data bases. We acknowledge discussion with Tony Williams, Christian Fischer and Patrick Bowman. H.N. is supported by the JSPS grant in aid of scientific research in priority area No.13135210.
This work is supported by the KEK supercomputing project 05-128.

\end{document}